# Semantic Segmentation of Anaemic RBCs Using Multilevel Deep Convolutional Encoder-Decoder Network

MUHAMMAD HAHZAD[1],  ARIF IQBAL MAR[1],  YED HAMAD SHIRAZI[1], AND ISRAR AHMED SHAIKH[2]

[1]Department of Information Technology, Hazara University Mansehra, Dhodial 21300, Pakistan
[2]Shaukat Khanum Memorial Cancer Hospital and Research Centre, Lahore 54770, Pakistan

Corresponding author: Arif Iqbal Umar (drarif.hu@gmail.com)

**ABSTRACT** Pixel-level analysis of blood images plays a pivotal role in diagnosing blood-related diseases, especially Anaemia. These analyses mainly rely on an accurate diagnosis of morphological deformities like shape, size, and precise pixel counting. In traditional segmentation approaches, instance or objectbased approaches have been adopted that are not feasible for pixel-level analysis. The convolutional neural network (CNN) model required a large dataset with detailed pixel-level information for the semantic segmentation of red blood cells in the deep learning domain. In current research work, we address these problems by proposing a multi-level deep convolutional encoder-decoder network along with two stateof-the-art healthy and Anaemic-RBC datasets. The proposed multi-level CNN model preserved pixellevel semantic information extracted in one layer and then passed to the next layer to choose relevant features. This phenomenon helps to precise pixel-level counting of healthy and anaemic-RBC elements along with morphological analysis. For experimental purposes, we proposed two state-of-the-art RBC datasets, i.e., Healthy-RBCs and Anaemic-RBCs dataset. Each dataset contains 1000 images, ground truth masks, relevant, complete blood count (CBC), and morphology reports for performance evaluation. The proposed model results were evaluated using crossmatch analysis with ground truth mask by finding IoU, individual training, validation, testing accuracies, and global accuracies using a 05-fold training procedure. This model got training, validation, and testing accuracies as 0.9856, 0.9760, and 0.9720 on the Healthy-RBC dataset and 0.9736, 0.9696, and 0.9591 on an Anaemic-RBC dataset. The IoU and BFScore of the proposed model were 0.9311, 0.9138, and 0.9032, 0.8978 on healthy and anaemic datasets, respectively.

**INDEXTERMS** CNN, deep learning, blood cells, segmentation analysis, semantic segmentation, multi-level deep convolutional encoder-decoder network.

## I. INTRODUCTION

With the quick expansion of CNN  architecture, deep learning has attained a great milestone in computer vision. The development of AlexNet [1] architecture brings the convolutional neural network (CNN) to a dominant position within the domain of computer vision. After that, several neural networks based on CNNs were developed one after another. Out of them, the most important ones are ResNet [2], DenseNet [3], GoogLeNet [4], CliqueNet [5], VGG-Net [6], ResNeXt [7], Xception [8] Inception-V4 [9] and NasNet [10] etc. all these outperform several times during ImageNet

The associate editor coordinating the review of this manuscript and approving it for publication was Zhouyang Ren .

competition. Due to the deep internal framework of these models, CNNs have the upper hand concerning accuracy, performance and are sometimes more efficient than human observation in some areas. Recently, CNNs have been widely used in the three core domains of digital image processing, i.e., object detection, image segmentation (Semantic Segmentation), and classification [11]–[13]. Image segmentation refers to identifying the region of interest (ROI) within an image. With promising results, various traditional segmentation methods [14]–[20] have been used in the medical image domain, especially for blood cell segmentation. But they failed to analyze the blood image at pixel level and ignored the semantics of ROI, especially in Anaemic red blood cells (RBCs.)

Anaemia is a condition in which a significant change in shape, size, and count of RBCs mass occurs [21]. Several techniques have been published to identify Anaemia in RBCs but ignore the pixel-level deep machine learning analysis. However, diagnosing blood-related disease greatly assisted with cytogenetic, molecular, and immunological tests. But, physical and quantitative descriptors like morphological, texture, and color play a crucial role in the diagnostic purpose of several morphological blood-related diseases like anaemia [22]. Recent research in semantic segmentation ignores the blood cell analysis, especially segmentation and pixel-level classification of abnormal RBC morphology due to Anaemia [39]–[45]. Red blood cells mostly show relatively the same geometry for shape and size along with overlapped structure. Most of the prior approaches do not consider the overlapped structure of RBCs and perform segmentation only on a single cell at the region level, not pixel-level. Segmentation of blood cells considering overlapping structures is still challenging due to the ambiguous, complex, and homologous assembly of cells. Current research also ignores the ground truth for crossmatching analysis and evaluation of their architecture. Precise identification of pixel-level information regarding the shape and size of infected red blood cells to normal is still challenging. Detection and treatment of blood-related disease at an earlier stage may overcome the worst results of the condition. Several recent research works [42], [46]–[48] used U-Net, Seg-Net, and FCN-AlexNet for blood cell segmentation. However, the issues regarding the prediction of change in morphological features of RBCs, like changes in shape and size of infected cells, remain to be solved.

For pixel-level segmentation, semantic segmentation gives outstanding results along with preserving semantic information. The fundamental concept of semantic segmentation is assigning distinct labels to each pixel of an image which is a challenging task in computer vision [14]. Due to the provision of pixel-level segmentation and classification, several real-world application got benefits from this technique i.e. video object segmentation [15]–[17], pedestrian detection [18]–[20], [23], [24], defect detection[25], therapy planning [26], [27], medical image diagnosis [28]–[30], computeraided diagnosis [31], [32] and other computer vision fields [33]–[36]. The pixel-level semantic information assists the intelligent systems in grasping spatial positions for making important decisions. With the rapid advent of deep learning approaches, semantic segmentation primarily depended on deep learning. Concerning the characteristics and different processing granularity, deep learning-based semantic segmentation is divided into two categories, i.e., 1) Semantic segmentation based on pixel classification and 2) Regional classification. Semantic segmentation based on pixel classification extracts semantic information and features from many labeled images. Using this information, it labeled and classified each pixel during end-to-end training to achieve semantic segmentation.

In contrast, semantic segmentation based on regional classification clubs the classifier with traditional image processing approaches for segmentation and classification purpose. Traditional methods segment the ROI within an image, and then semantic classification is carried out using a classifier. Several semantic segmentation approaches use CNNs for the assignment of distinct labels to each pixel [5], [9], [37], [38].

This research proposed a multi-level deep convolutional encoder-decoder (DCED-Net) network for semantic segmentation of blood cells to overcome the above problems. We also developed two state-of-the-art blood cell datasets that will help the research community for the testing of semantic segmentation algorithms along with ground truth for crossmatch analysis. For experimental purposes, we have collected a total of 2000 annotated blood images from Shaukat Khanum Memorial Cancer Hospital and Research Centre Lahore (SKMCH&RC) under the supervision of an expert pathologist. This dataset comprises 1000 normal and 1000 anaemic RBCs images along with CBC and morphology reports. Each image is further divided into manually generated ground truth under the supervision of a pathologist for the authentication of segmented results.

The key contribution of this research work is as follows:

1. First, we developed a multi-level deep convolutionalencoder-decoder (DCED-Net) network for semantic segmentation with forward and reversed connections between two consecutive levels to exchange semantic information.

2. We find the semantic segmentation of normal andanaemic RBCs with accurate global and class-wise pixel accuracy that helps to diagnose disease.

3. Development of two state-of-the-art Anaemic RBCsdatasets comprises 2000 images, including CBC and morphology reports of each image.

4. This dataset is also equipped with 2000 manually generated ground truth for each image that authenticates the performance of the proposed architecture using crossmatch analysis.

The rest of the paper is organized as follows: Section II comprises related work. Section III describes the proposed methodology and data collection/preparation of the current work. Experimental work, results and discussion, conclusion and future work are described in sections IV-VII.

## II. RELATED WORK

Automated blood cell segmentation plays a vital role in quantitative analysis in biomedical image processing [40]. Cell segmentation approaches are divided into supervised and unsupervised categories. Un-supervised segmentation includes classical approaches like watershed transformation [49], region growing, gradient flow [50], semi-automated methods [51], etc. While fully convolutional networks (FCN) and CNN have been used in supervised segmentation methods. In [40], authors developed a deep semantic segmentationbased architecture for the classification of RBCs. They enhanced the classic U-net architecture by adding deformable convolutional layers to correct sickle cell disease's red blood cells classification. But this model needs more work for the correct identification of cell counts in overlapped areas. Another work has been done by [26] using a sliding window approach for aggregate labelling of each pixel in the blood cell. But this work reduces the resolution of segmentation. The authors [52] used an online regionbased segmentation method to detect cancerous red blood cells in the BCCD dataset. This study helps the radiologist for the detection of cancerous cells but is not able to analyze them at the pixel level. For the maximization of semantic information, [38] introduces a loop-based c with an encoder-decoder structure. They created a loop between two layers by adding reverse and forward connections. Each layer works in both directions, i.e., forward and reverse direction. Due to the clique block structure, the calculation cost increases as the number of layers increases. The total depth of the network was 194 layers with encoder (82 layers) and decoder (112 layers). This network outperforms at CamVid and Cityscapes. Dynamic attention networks have been developed by [48] for semantic segmentation of general-purpose images using PASCAL VOC 2012 [52] and Cityscapes [53] datasets. In [48] author developed a network with two modules: Deformable attention pyramid (DAP) and Fusing Attention Interim (FAI) modules to perform self-adjustable descriptions of high-level output and back-propagation guidance of information. Semantic features of images were extracted by using Res-Net 101 [2] and DAP. Fusing Attention Interim used global average pooling and feature fusion into the block to build the smooth connection between decoder and encoder.

Another semantic segmentation network has been designed by [54] to recognize the complex Chinese character. They combined three modules, i.e., FCN-ResNet50, Wubi-like label coding, and CRF, to develop LCSegNet. For pixellevel classification, FCN-ResNet50 [55] was used, Wubi-like label coding was used to convert Chinese characters into 140-bits labels. In contrast, the CRF module overcame invalid code labeling for the transformation matrix among different groups. ResUNet [56] framework comprises Dice loss function along with hybrid UNet encoder-decoder architecture. The author used the dice loss function and pyramid scene parsing pooling, residual connection, multi-tasking inference, and atrous convolution to account for the precise semantics of remotely sensed data. The experiment was performed on the ISPRS 2D Postsdam dataset. Selection loss and attention loss were used by [57] with noisy annotation for semantic segmentation of the PASCAL VOC 2012 dataset. This architecture works in three steps: First, it generates initial proxy annotation. In the second step, the mask coring strategy was used for the selection of high-quality annotation. The Selection and Attention (SAL) network was applied to retrain the segmentation network in the last step. For the quantitative measurement of edge segmentation, [58] proposed a regionbased evaluation index architecture. This work aimed to enhance the segmentation at the region of interest (ROI) edges within an image by using two losses, i.e., los Lce and loss Lhp. Task decomposition technique has been used by [59] for semantic segmentation of Robotics scene, Brain tumour, and Retinal fundus images. To reduce the distance between the results of pixel-level semantic segmentation and the instance class prediction tasks, they used a novel sync-regularization multi-step approach. The substructure in the satellite image dataset was addressed by [62] using a cross-channel structure and large kernel size. To boosts the image features, they used the GAN network with a feature attention mechanism. They got mIOU with 88.15% at PASCAL VOC 2012 dataset. Semantic segmentation was also used as transfer learning [60] between various deep learning classifiers to identify different types of crops. The author used a shortened version of the SegNet framework, known as SegNet-Basic. A hybrid version of SegNet known as Random-Walk-SegNet (RWSNet) was developed by [61] for enhancing blurry object boundaries in remote sensing images. Undirected graph, gradient, probability map of SegNet, and random walk algorithm were used with basic SegNet to develop the RWSNet framework. Several recent research works [42], [46]–[48] used U-Net, Seg-Net, and FCN-AlexNet for blood cell segmentation. However, the prediction of percentage change in morphological blood cell features like change in shape and size of infected cells remains to be solved.

Previously developed blood cell datasets like ALL-IDB-I, ALL-IDB-II [65], extended ALL-IDB [66], BCCD, IUMSIDB [67], SMC-IDB [68], BS_DB3 [69], Ash bank, BBBC [70] dataset equipped with small no of images with little RBCs elements as shown in Table 2. ALL-ID-I contains only 108 images with just 39000 blood elements. All 108 images have no masks of authentication of proposed results. Other datasets also lack detailed information regarding manual ground truth for authentication purposes and blood cells elements. SMC-IDB contains only 367 WBC-type images without any ground truth. IUMS-IDB and malarial datasets comprise 196 (WBCs) and 848 (RBCs) images, respectively. Approximately all datasets have no ground truth fortheauthenticationofproposedtechniquesexceptextended ALL-IDB. No dataset targets the morphological changes in RBCs due to anaemia disease.

## III. METHODOLOGY

The workflow chart of the proposed semantic segmentation architecture is shown in Figure 1. This architecture is mainly divided into the following steps:

- Image data collection and Pre-processing
- Network Architecture
- Segmentation

### A. IMAGE DATA COLLECTION AND PRE-PROCESSING

The proposed work randomly selected a total of 100 participants for blood image collection. Out of which, 50 were healthy, and 50 were anaemic patients. Data collection was carried out under the supervision of an expert pathologist from Shaukat Khanum Memorial Cancer Hospital and pixels resolution) were collected along with relevant CBC and morphology reports of each image. The ground truth of each image was also prepared under the supervision of an expert pathologist to authenticate generated output. After collecting the RBC dataset, all the images were categorized into two main groups based on their cell type: Group I: Healthy RBCs image and Group II: Anaemic RBCs.

Thequalityofbloodimagesandcellmorphologyregarding healthy and infected cells was critically examined by a hematologist before the image processing. This critical examination helps us for the selection of anaemic and healthy images. Each image contains approximately 83 RBCs elements. Therefore, our dataset comprises 4000 images (2000 original, 2000 ground truth) and about 1,992,000 RBC elements. These 1,992,000 RBC elements include normal microcytes, macrocytes, elliptocytes, and target RBCs. Samples images from healthy and anaemic RBCs are shown in Figure 2. The pre-processing step involves three primary

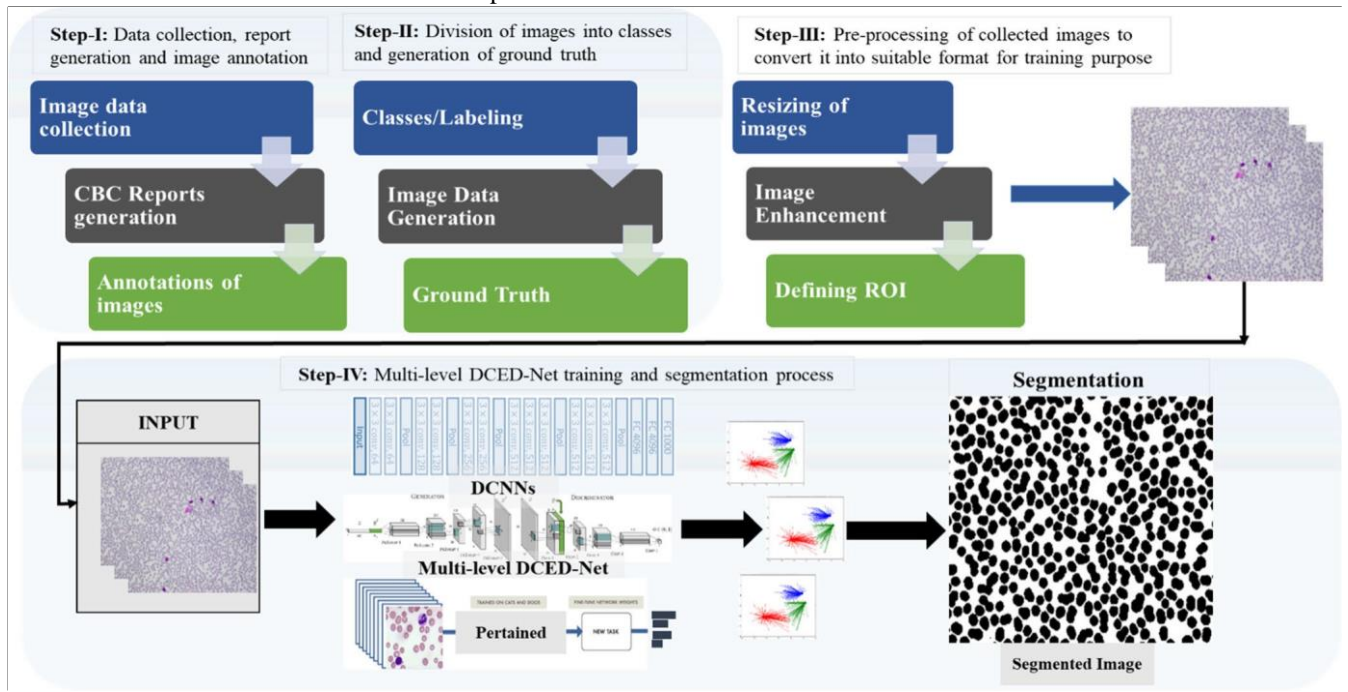

**FIGURE 1.** Flow chart of multilevel deep convolutional encoder-decoder network (DCED-Net). Step-I: The first step included blood image collection from Shaukat khanum hospital, Lahore, CBC and morphology reports generation of each image and image annotation abut pathologist decision. Step-II: This step involves dividing these images into classes, i.e., healthy or anaemic image and ground truth generation of each image for evaluation purposes. Step-III: In a pre-processing step, image resizing, enhancement and definition of region of interest (ROI) are carried out. Step-IV: In this step, training of DCED-Net is carried out that produces a segmented image as output.

Research Centre (SKMCH&RC) Pakistan. Olympus Dp27 8.9-megapixel CMOS sensor captured RBC images with a 4K resolution of microscopic digital camera. Twenty blood smear samples were obtained from each of the 100 participants. Using these samples, we have prepared 2,000 blood smear slides from the routine workload of the hospital pathology lab. A total of 2,000 digital images (490 × 480 activities, i.e., first: grayscale conversion of RGB images, second: pixel fusing, and then unity mask generation. Contrast normalizationwasusedtoreducenoisy,blurrypatternstoproduc esharp edges with high contrast images. Resizing of each image is also done according to the underlying model, i.e., 320 × 320 × 3. Pre-processing phase comprises the following steps, also shown in Figure 3:

- Algorithm read both type of images (original and finetuned manual generated masks) from memory.

- All the images were checked for three channels of color, and pixel labelling fusing was also initiated in this step.
- Wiener and Laplace's filters were applied in spatial domains to separate the region of interest (ROI) from the background.
- A pixel ID, PID = 0, was assigned to all pixels of ROI to generate the unity mask, and PID = 1 was given to the image background.
- All the pixels of the RBC element and background are pointed out individually.
- After that, segmented ROI (RBC element) was generated with pixel value zero.
- After that, each image was resized into 320 × 320 × 3 resolution and passed to the model for training.

### B. NETWORK ARCHITECTURE

Clique Net inspires a Multi-level deep convolutional encoderdecoder network's design. DCED-Network is arranged to maximize the transmission of semantic information and best optimum features selection from one level to another. Each level of DCED-Net comprises the encoding and decoding phase.

### 1) ENCODING

During pre-processing step, all the images are resized into 320×320×3 resolution. These 320×320×3 images are given as input to the encoding block of the DCED-Net for extracting features to encode them into optimum features. The encoding block comprises five pools of layers. Each pool comprised convolution layers with stride size 3, batch normalization, and average pooling layers. The internal operation of each layer's pool is shown in Figure 4. After convolution operation, the output is concatenated with the original input and given to the next pool during encoding and decoding phases. The dropout was added only on the last two pools to reduce and prevent the selection of irrelevant features. The first pool receives input images with size 320×320×3 and reduces them into 160× 160 × 32 by applying 3 × 3 convolutions and 2 × 2 average pooling, as shown in Figure 5. On the 2nd and 3rd pools, the input image reduces upto 40 × 40 × 128. On the 4th and 5th pools, additional features and attributes are extracted from the image to reduce them into 10 × 10 × 512. At the end of the encoding block, the input image obtained with 10×10×512 size contains the best optimum feature for decoding purposes. The encoder performs a downsampling operation, as shown in Figure 5.

### 2) DECODING

The decoding block performs transpose convolution (TC) on the output of the encoding block to generate a segmented image from 10 × 10 × 512 to 320 × 320 × 3. It consists of 5 pools of layers. Each pool consists of the same layer as the encoding block. Each pool in the decoding block is connected with the same pool in the encoding block using skip connection to preserve the semantics of the image, as shown in figure 4. A dropout of 0.2 was applied on the initial two pools to maintain the original texture and features of an image and reduce overfitting. Every decoding pool received two inputs: 1) One from the previous pool and 2) The second is from the corresponding same-sized encoding pool to improve the feature selection and texture preservation. The main operation of the decoding block is the upsampling of a low-resolution image to a high-resolution segmented image. The decoding block retrieves the high-resolution image by applying 3 × 3 transpose convolution with max pooling. Each level in DCED-Net receives input in image format and produces output in the segmented image. So the input from one level to the other is the image and also ground truth.

### C. SEGMENTATION

Segmentation is the most challenging and critical task for digital blood image analysis. This is because of the complex and crowded structure of the blood cell elements. Accurate segmentation greatly assists in subsequent levels of classification. For accurate and efficient semantic segmentation of anaemic RBCs, we have used a DCED-Net along with pre-processing steps. Here is the pseudo-code for our DCEDNet model and evaluation matrices used for performance measures.

**BEGIN**
**# Two types of data will be loaded i.e. Anaemic Red Blood Cells, ii. Healthy Red Blood Cells** def Load_Data:

```
    For each Image in AnemicImageSet:
        AnemicRBC[Image] = List of AnemicRBCImages
        AnemicLabelCount = AnemicLabelCount + 1
    Next
    For each Image in HealthyImageSet:
        HealthyRBC[Image] = List of HealthyRBCImages
        HealthyLabelCount = HealthyLabelCount + 1
    Next
    If        data_loaded_successful
        Return True
    Else
        Return False
Return data_loading_status
```
**# Pre-Processing Data Sets**
```
While ImageCount <= AnemicLabelCount
    If(Typeof(AnemicRBC[Image])=''RGB'' than
        AnemicRBC[Image]) = ConvtoGrayScaleImg
        (AnemicRBC[Image]))
        Else If(PixelLabelingMaskFusion == True)than
            PixelValue = 0
            Generate UnityMask
        End If
```

End While
While ImageCount <= HealthyLabelCount
   If (Typeof (AnemicRBC[Image]) = ''RGB'' than
     AnemicRBC[Image])= ConvtoGreyScaleImg
     (AnemicRBC[Image]))
   Else If (PixelLabelingMaskFusion == True) than
     PixelValue = 0
     Generate UnityMask
   End If
End While
**# Generate Image with Segmented ROI having Zero-Pixel Status**
While    ImagePixel(AnemicRBC[Image])   =   0   OR ImagePixel(HealthyRBC[Image]) = 0
 If Background = 1 and RBCElemnet = 0 than
   Set MaskPixelID = UNIQUEPIXEL(RBC, Background)
     Resize MaskedImageWidth = 320
     Resize MaskedImageHeight = 320
  SetOriginalPixelID=UNIQUEPIXEL(RBC,Background)
     Resize OriginalImageWidth = 320
     Resize OriginalImageHeight = 320
 End If
 SaveFile ''MaskedDataSet''
 SaveFile    ''OriginalDataSet''
 End While
**# FEATURES EXTRACTION**
While Not All Data Loaded

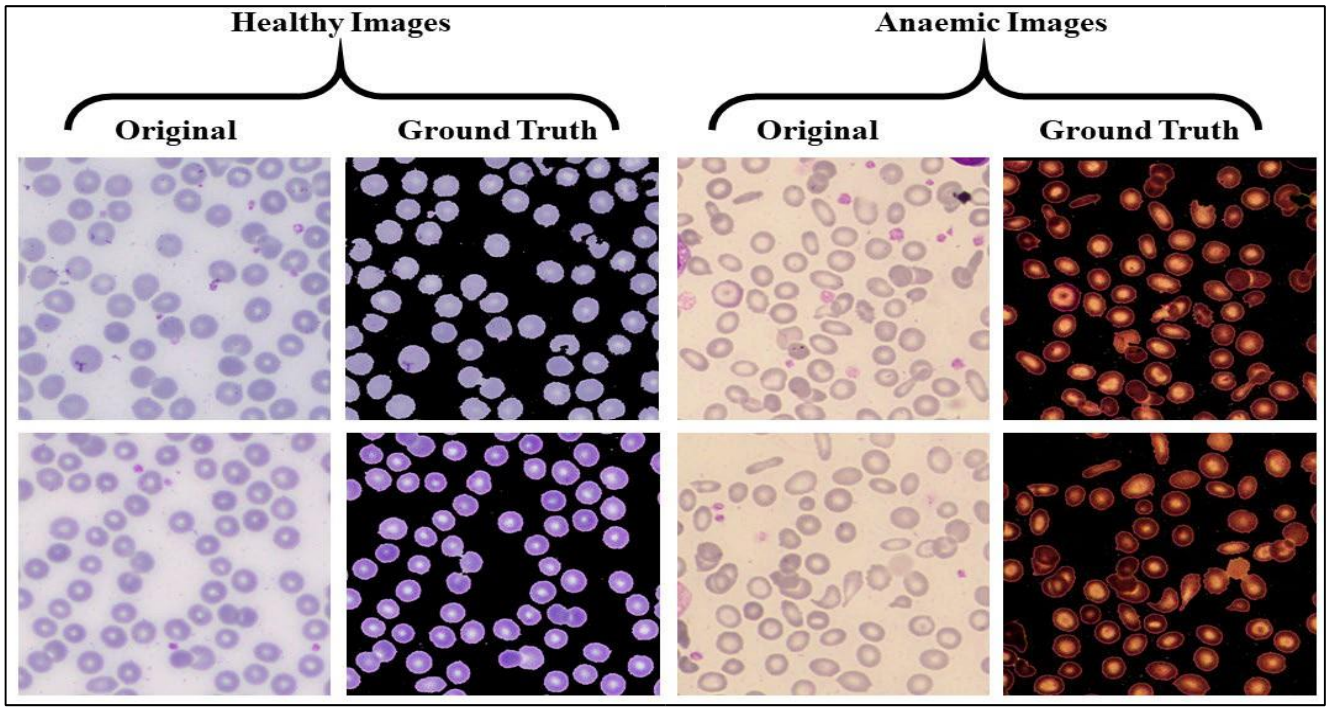

**FIGURE 2.** This diagram shows the samples from healthy and anaemic images from the proposed dataset. The images of this dataset are captured under various illumination as shown in the diagram.

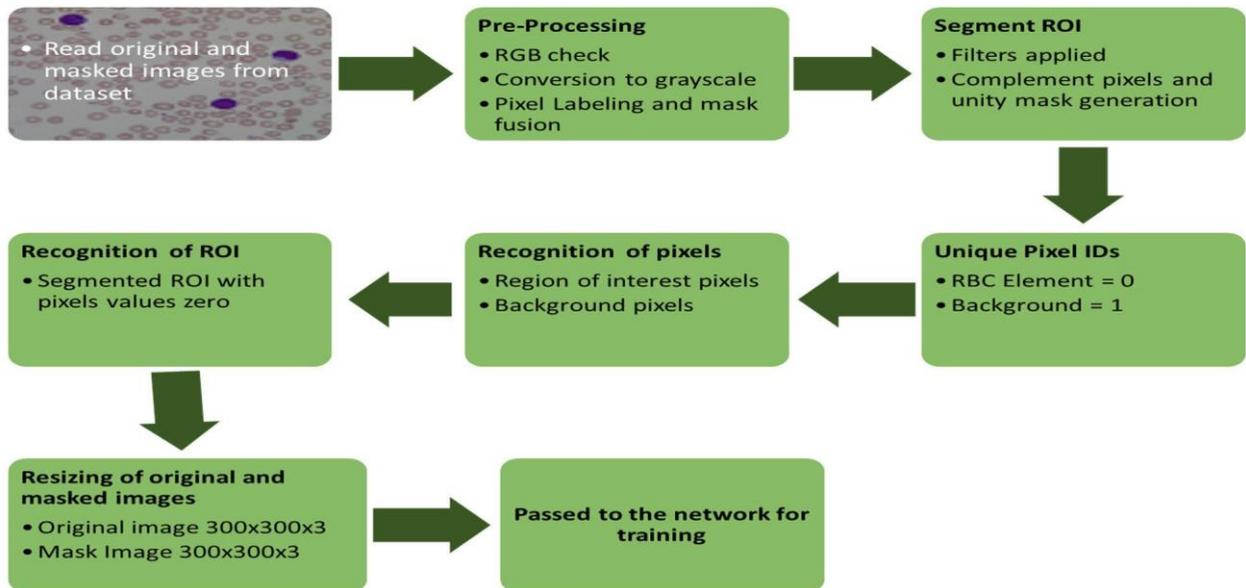

**FIGURE 3.** All pre-processing steps of DCED-Net.

For each Image in MaskedDataSet:
        Extract(ListofFeaturesMaskedData)
Next
For each Image in OriginalDataSet:
     Extract(ListofFeaturesOrginalDataSet)
Next
TrainingDataSet = 80% of Extracted Features
TestDataSet = 20% of Extracted Features

End While
**#LOADING DCED-NET and Configuring Training Parameters**
While DCED-NET Loaded
    **Configure** Batch_Size_minimum ← 2 **Configure**
    Total Epoch ← 300-500
    **Configure** Iterations per epoch ← 150
    **Configure** Learning rate ← 1e-4

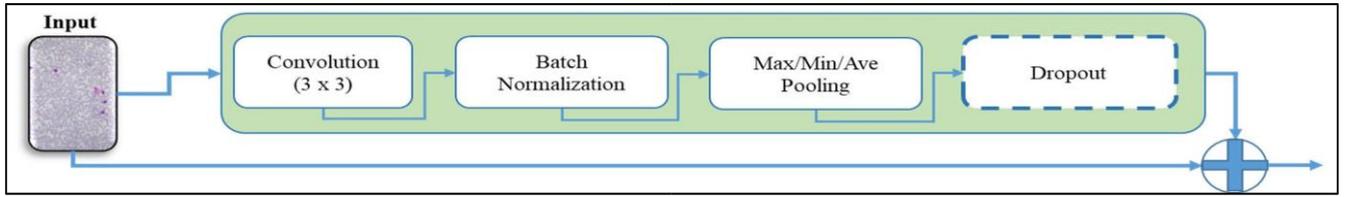

**FIGURE 4.** The internal operation of each layer's pool in DCED-Network.

$$\acute{N} = \acute{L}_1 + \acute{L}_2 + \acute{L}_3 + \acute{L}_4 \tag{1}$$

Procedure TrainModel
    For I = 1 to TotalEpoch Randomly
        sample the training data with
        replacement
        Count Healthy and Anaemic Pixels of RBC
        Create separate Segments for each Healthy and
        Anemic Pixels
    End For
  End Procedure
End  While  #   **Compute**
**Evaluation Metrics**

Calculate Accuracy $\leftarrow \frac{TP+TN}{TP+TN+FP+FN}$

Calculate IoU $\leftarrow \frac{Target \cap Prediction}{Target \cup Prediction}$

Calculate BF Score $\leftarrow 2(\frac{Precision*Recall}{Precision+Recall})$

**END**

Above pseudo-code is the detailed description of the segmentation technique. For understanding, we have divided the segmentation process into 05 primary steps: 1) In the first step, the system uploads original and manually generated images for pre-processing. 2) During pre-processing, unity mask generation, pixel fusing, and implementation of morphological operation were performed. Resizing and removing noisy and blurry patterns of an image were also handled during this phase. 3) In the 3rd step, all the input data is divided into test and train images. 4) Proposed DCED-Net was loaded for training purposes. Network architecture diagram of DCED-Net are shown in Figure 5. 5) In the last step, segmented healthy and Anaemic RBCs were generated along with evaluation matrices measurement. A brief description of semantic segmentation is given in Figure 6.

Inspired by LSTM and recurrent network structure, we developed a new multi-level deep convolutional encoderdecoder network. Unlike prior CNN structures, we divided the encoder-decoder network into different levels with intralevel, inter-level forward, and reverse connections between any two consecutive layers and levels/groups, respectively, as shown in Figure 6. We address DCED-Net as $\acute{N}$. The network $\acute{N}$ is a composition of multiple levels of the encoderdecoder networks. For simplicity, each level is represented as $\acute{L}$. The DCED-Net is defined in equation 1.

The last level of architecture is also connected with all previous tiers. The final layer of each pool is equipped with two variables, i.e., 1) Threshold value ($T_o$), and 2) Comparison function ($C_o$). The value of $T_o$ is set as 50%, 80%, and 95% for levels 1,2, and 3 of DCED-Net, respectively. These values are set as a condition along with training parameters at the start of the training. Now, each level received two inputs: one actual ground truth (G) and the second is the previous level's output. Every encoder pool extracts a feature map fromtheinputimageandgeneratesasegmentedimageduring decoding phases. At the last layer of the decoding phase, comparison function $C_o$ compares the extracted features map (F) and generated image ($\mathcal{S}_n$) of level $n$ with the $T_o$ value of $n^{th}$ level ( ) and ground truth (G) respectively. Here $n$ represents the current level of the network. If the $C_o$ function got a value greater than or equal to the current level's threshold, the $\mathcal{S}_n$ andpassed as input to the $n + 1$ convolutional level for further processing.

On the other hand, if the value of $C_o$ is less than $T_o$, the output of the $n^{th}$ convolutional level is passed as input again to the same level for more optimal feature selection. In this way, the best optimum feature selection is achieved, preserving pixel-level semantic information at each layer and architecture level. The whole comparison function is formulated in the following three equations.

$$C_o = \mathsf{F}_n \equiv T_o \,\& \, \mathcal{S}_n \equiv G \tag{2}$$

$$C_o \geq T_o \text{ then output} \rightarrow n + 1 \; level \tag{3}$$

$$C_o! \geq T_o \text{ then output} \rightarrow n^{th} \; level \tag{4}$$

Here $\equiv$ operator act as a comparison operator. The output of the last level compares with ground truth data. If the system achieves the required accuracy, it evaluates statistical matrices for performance evaluation. Otherwise, the last level output is sent back to the appropriate level for the re-selection of features. Each layer and level within the architecture act as input and output, which means they are more densely connected than prior networks [3], [5]. This strategy enhances the pixel-level feature selection and preservation of semantic information of ROI. The proposed Multi-level DCED-Net has some unique properties. A deep look at the architecture shown in Figure 6 explains that the model is more densely feature-demanding. With given block of $n$ layers, CliqueNet [5] need $A^2_n$ (Permutation operator) group of feature for single clique block while DenseNet [3] required $C_n^2$ (Combination operator), but we employed $A$ $\frac{2}{n} + A_{n+1}^2 + A_{n+2}^2 + A_{n+3}^2 + \ldots A_{n+n}^2 \big)$ for each level as well as for every layer within each level. This shown at the bottom of the next page. Here, $*$ denotes convolutional operations with feature parameters $W$. while $Y$ means that DCED-Net needs product of permutation features. The training process formulated in equation 5, as is the non-linear activation function. Parameters weights $W_{ij}$, keeps the pixel features for comparison operations and future use. In DensNet, filters increase linearly as the depth of the net increases [62], which causes rapid growth of parameters, that may feed irrelevant features. While in cliqueNet, every blockprovidedtheirfeaturestothenextblockbeyondstage-II irrespective of block output accuracy. But, in the proposed model, the output of each level compares with the previous level's using the $C_o$ function, and only those features are fed to the next level, which is optimal. It turns out that this is a more optimized feature selection way. Densely connected DCED-Net has a solid ability to preserve pixel-level semantic information with the help of recurrent structure at each layer and level with a feedback mechanism. Forward and feedback connections are densely connected at each level to maximize the best feature map and information flow. Due to this structure, our model outperforms for reducing background noises, paying more concentration only on ROI, and achieving competitive performance over existing techniques. For the evaluation and authentication of proposed model results, we used accuracy, IoU, and BF score. Crosstrain analysis was also performed with manually segmented ground truth under the supervision of expert pathologists.

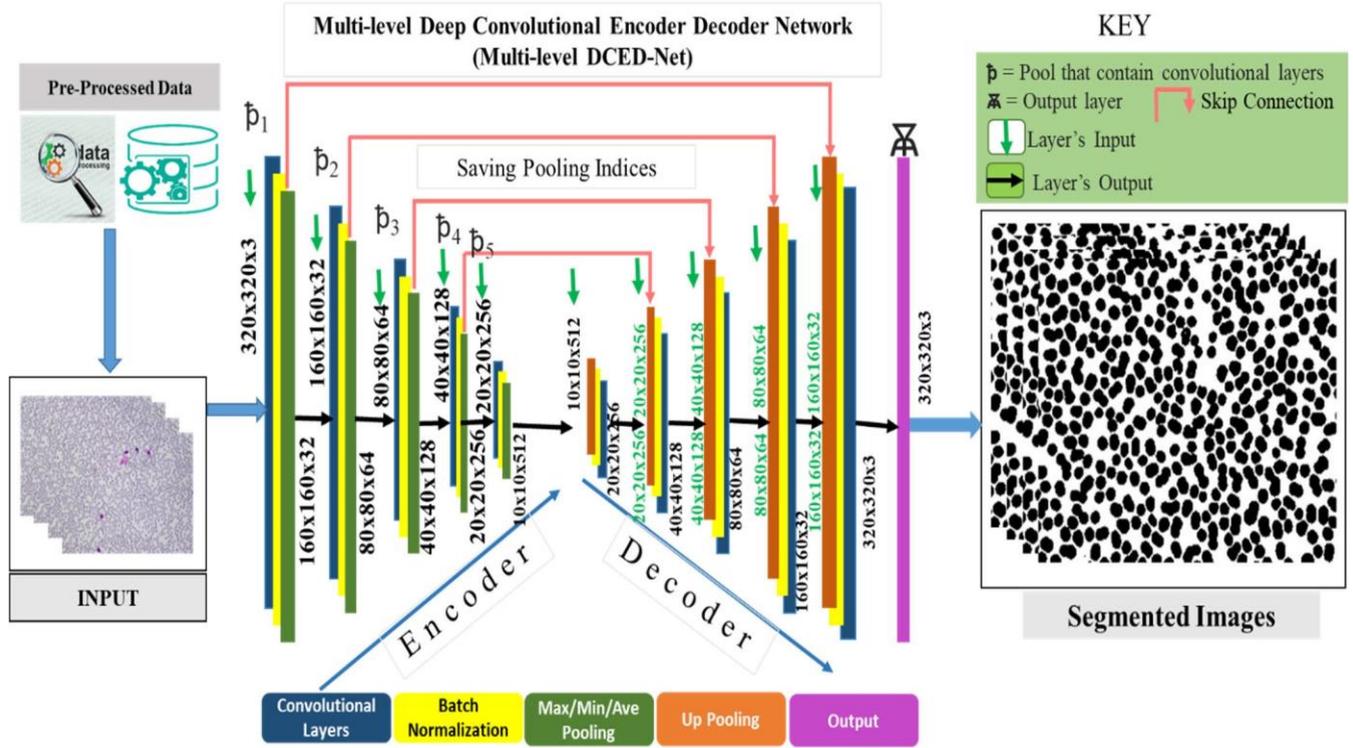

**FIGURE 5.** Details of layer's input and output of single network used in multilevel deep convolutional encoder-decoder network (DCED-Net) for semantic segmentation.

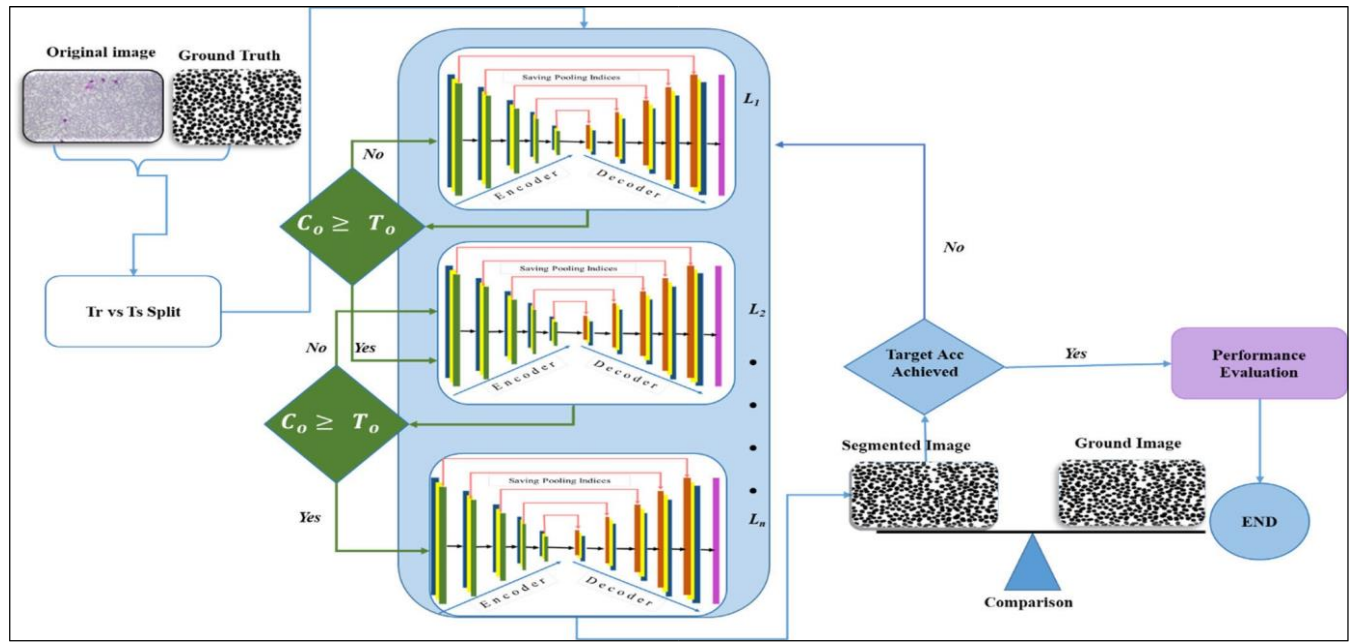

**FIGURE 6.** Synaptic view of semantic segmentation process in densely connected DCED-Net. Co represents the comparison function, To represent the predefined threshold value., Tr means train dataset, Ts mean test dataset.

### 1) LOSS FUNCTION
The loss function of the network is adjusted using the mean square error (MSE) technique. MSE is the average of the squares of the errors between predicted image ($p$) and ground truth ($g$). It is calculated by equation 6.

$$MSE = \frac{1}{n}\sum_{i=1}^{n}(g_i - p_i)^2 \qquad (6)$$

MSE of the proposed model is measured at each network level, and feedback is given to the network for further training. MSE is measured continuously until or unless the network got a value greater than the predefined threshold value $T_o$. As the network got desired MSE value, the training of the current level was stopped, and output was given to the next level for further processing with an updated value of $T_o$.

## IV. RESULT AND EVALUATION MATRICES
### A. TRAINING AND TESTING
The proposed semantic segmentation architecture is trained and tested on two datasets, i.e., Healthy-RBC and AnaemicRBC. A total of 2,000 (1000 healthy and 1000 anaemic) peripheral blood smear images were collected from 100 individuals, i.e., 20 images from each with approximately 1,992,000 (Original and Ground truth) RBCs elements. We have also generated 2,000 manually segmented ground truth for authentication purposes.

Figure 2 demonstrates the category distribution and dataset organization. Out of 2,000 images, the system randomly chooses 1800 images for training and 200 images for testing.

The proposed model was trained on 400 epochs, 150 iterations per epoch on both datasets with an initial learning rate of $1e^{-4}$ and a minibatch size of 2. Model execution carried out on GEFORCE RTX-3060 GPU with 24 GB of RAM.

### B. EVALUATION
All the results of the underlying model were evaluated based on two evaluation projections. 1) Iteration vs Accuracy and Iteration vs Loss Projection, 2) Epochs vs Accuracy and Epochs vs Loss Projection.

### 1) ITERATION VS ACCURACY AND ITERATION VS LOSS PROJECTION
The DCED-Net was trained on 400 epochs. The training procedure comprises 60,000 iterations with 150 iterations per epoch. In Figure 7 part (A, B) shows the loss and accuracy projection against the iterations on Healthy-RBC and Anaemic-RBC datasets, respectively. The green line demonstrates the ratio between min-batch accuracy and iteration, while the blue line explains the relation between min-batch loss with each iteration. On the Healthy-RBC dataset, with the start of the training procedure, the accuracy graph starts from 0 and directly jumped into 0.6689 with 0.8096 loss on completion of 150 epochs. But, the Anaemic-RBC dataset shows relatively less accuracy, i.e., 0.6577 with 0.8707 loss. After completing 162 epochs with 24,300 iterations, the accuracy graph goes to 0.9731 with downfall of the loss graph to

0.0634 from 0.8096 on Healthy-Dataset. While in Anaemicdataset, after 24,300 iterations, the accuracy graph goes up to 0.9556 with 0.1266 loss. As the iteration goes to 45,000, the spike of the accuracy graph touches the highest point, i.e., 0.9856 and 0.9736 on healthy and anaemic datasets, respectively. While after 52,000 iterations, the model loss goes down to the lowest value, i.e., 0.0252 on the healthy dataset and 0.0351 on the anaemic dataset.

### 2) EPOCHS VS ACCURACY AND EPOCHS VS LOSS PROJECTION
In Figure 7 part (C) the green line explains the relationship between accuracy achieved against each epoch. The blue line explains loss during training on the healthy-RBC dataset. While part (D) shows the ratio between epoch vs loss and accuracy on the anaemic-RBC dataset. On the Healthy-RBC dataset, at the first epoch, the training accuracy was 0.6689 with 0.8096 loss. As the epochs increase to 100, the accuracy goes upward and touches 0.9707, decreasing the corresponding loss down to 0.0789. After that point, the accuracy and loss fluctuate between approximately 0.9754 to 0.9856 and 0.0317 to 0.0215 on both datasets, respectively. The highest accuracy is achieved at epoch no

$$Z_{on} = \left( f_i \left( \sum_{i<_j} W_{ij} * Y_i^n + \sum_{x>j} W_{xj} * Y_x^{n-1} \right) + f_{i+1} \left( \sum_{i+1<j+1} W_{ij+1} * Y_{i+1}^n + \sum_{x+1>j+1}{}_{n-1} W_{xj}{}^{+1} * Y_{x+1}^{n-1} \right) \cdots \right)_? \\ \left( f_{i+n} \left( \sum_{i+n<j+}{}_n W_{ij}{}^{+n} * Y_{i+n}^n + \sum_{x+n>j+}{}_n W_{xj}{}^{+n} * Y_{x+n} \right) \right) \quad (5)$$

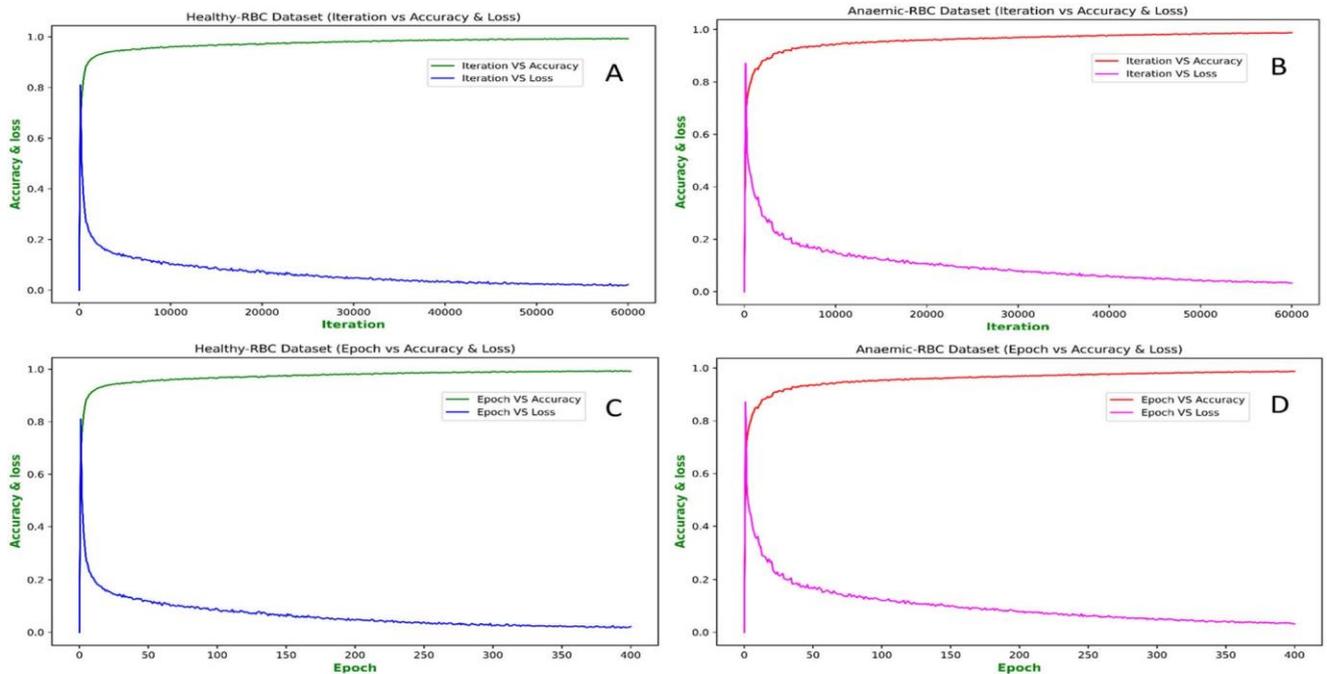

**FIGURE 7.** Part (A) Iteration vs accuracy and loss projection on the healthy-RBC dataset, Part (B) Iteration vs accuracy and loss projection on Anaemic-RBC dataset, Part (C) Epoch vs accuracy and loss projection on Healthy-RBC dataset, Part (D) Epoch vs accuracy and loss projection on Anaemic-RBC dataset.

345, i.e., 0.9856 and 0.9736 on healthy and anaemic datasets, respectively. At the same time, minimum validation loss against epoch is achieved at epoch no 324 with 0.0171 on the healthy dataset and 0.0724 on the anaemic dataset.

### C. EVALUATION MATRICES

We have further evaluated our model on three fundamental statistical matrices, i.e., 1) Accuracy (mean and global); 2) MeanBFScor; and 3) Intersection over union (IoU) (weighted and mean)

#### 1) ACCURACY

In semantic segmentation, accuracy is used to determine correctly classified pixels of an image. Accuracy is the ratio between the product of correctly identified negative pixels (TN) and correctly identified positive pixels (TP) over the product of falsely identified positive (FP), falsely identified negative pixels (FN) along with TN and TP as shown in equation 6:

$$Accuracy = \frac{TN + TP}{TN + TP + FN + FP} \quad (7)$$

To find out the mean value of each accuracy, we have carried out a 5-fold training procedure. Table 1 shows the different accuracies of both RBCs datasets, i.e., healthy and anaemic, with global accuracy. Individual dataset training, validation, and testing accuracies with relevant losses are also shown in Table 1. Our model got 0.9856 and 0.9738 training accuracy for healthy and anaemic RBCs with 0.0253 and 0.0351 losses, respectively. While validation and test accuracies were 0.9760 and 0.9720 for healthy and anaemic datasets. Global training, test, and validation accuracies on both datasets simultaneously were 0.9796, 0.9728, and 0.9655, respectively, shown in Table 1

#### 2) MeanBFScore

The match between the predicted boundary of an image and the ground truth boundary is measured by BFScore matrix [63]. BFScore is calculated by taking the ratio of twice of recall and precision product over the sum of precision and recall as described in equation 7:

$$BFScore = \frac{2 * (Recall * Precision)}{Precision + Recall} \quad (8)$$

The BFScore of both datasets was calculated as 0.9138 and 0.8978 for the healthy and anaemic datasets, respectively shown in Table 1. At the same time, the overall MeanBFScore on both datasets was 0.9058, shown in Table 1.

### 3) INTERSECTION OVER UNION IoU

Intersection over Union (IoU) [64] is another class of statistical matrix used to evaluate semantic segmentation results. The primary function of IoU is quantifying the ratio between overlapping of predicted output and targeted ground truth. It is calculated by finding the balance between the intersection of target and predicted pixels overall masks and original image pixels. The mathematical form of IoU is shown in equation 8:

$$IoU = \frac{Target \cap Prediction}{Target \cup Prediction} \qquad (9)$$

Our model attained IoU of 0.9311with healthy-RBC while 0.9032 with Anaemic-RBC dataset, as shown in Table 1. Overall, MeanIoU was calculated as 0.9058 collectively on both datasets.

## V. RESULT AND DISCUSSION

The proposed model was trained, validated, and tested on two RBCs datasets, i.e., Healthy and Anaemic-RBCs. Both datasets were collected from SKMCH&RC Lahore. HealthyRBC datasets were collected from 50 patients with 20 samples from each. So, a total of 1000 healthy images were obtained. Likewise, 50 patients were selected randomly from the daily workload to collect anaemic-RBCs with 20 samples from each. So, a total of 1000 anaemic images were obtained. After the collection of images, ground truth was generated for both datasets to authenticate the predicted results. The proposed model is applied on both datasets individually and collectively to find the performance and accuracy. The model was implemented in two phases: 1) PreProcessing phase, manually generated masks, and original images were pre-processed and modified according to the model requirement. 2) In the second phase, a deep convolutional encoder-decoder network (DCED-Net) was initiated. DCED-Net comprises an encoder-decoder network and pre-trained VGG-16. The densely connected deep convolutional encoder-decoder framework is used to train and test both RBC-datasets, while VGG-16 was used for pixel-level feature extraction. Evaluation matrices like Accuracy, IoU, and BFScore were calculated to prove the performance of the model.

### A. DATASET

Previously developed blood cell datasets like ALL-IDB-I, ALL-IDB-II [65], extended ALL-ID [66], BCCD, IUMSIDB [67], SMC-IDB [68], BS_DB3 [69], Ash bank, BBBC [70] dataset equipped with small no of images with little RBCs elements as shown in Table 2. ALL-ID-I contains only 108 images with just 39000 blood elements. All 108 images have no masks of authentication of proposed results. Other datasets also lack detailed information regarding manual ground truth for authentication purposes and blood cells elements. While in this research work, first, we have developed two state-of-the-art blood cell datasets: 1) Healthy-RBCs dataset and 2) Anaemic-RBCs dataset with their relevant CBC and morphology report. These reports will help the research community to authenticate the proposed model results in respect of anaemia disease. Both datasets are equipped with 2000 images (1000 healthy, 1000 anaemic) with the ground truth of each image. So, our dataset repository comprises a total of 4000 images (2000 original and 2000 ground truth) with approximately 1,992,000 RBC elements. RBCs elements are categorized into four classes, i.e., normal, microcytes, macrocytes, elliptocytes, and target RBCs, as shown in Figure 8.

### B. ACCURACY

Table 1 shows the training, testing, and validation accuracy of both datasets. Our model outperforms on the healthyRBC dataset with training, validation, and test accuracies of 0.9856, 0.9760, and 0.9720, respectively, compared to the anaemic-RBC dataset. At the same time, BFScore and IoU of the proposed model on anaemic-RBCs images were 0.8978 and 0.9032, respectively. The BFScore and IoU on the healthy-RBC dataset were calculated as 0.9138 and 0.9311, as shown in Table 1. All these figures reveal that the proposed model outperformed the healthy-RBC dataset compared to the anaemic-RBC images. This is because anaemic-RBCs images are more complicated than healthy RBCs. We have carried out 05-fold training, validation, and testing of the underlying model and got global training, validation, and testing accuracies 0.9796, 0.9728, and 0.9655, respectively, as shown in Table 1.

### C. GLOBAL AND CLASS-WISE PIXEL COUNTING

Accurate blood element examination is beneficial for predicting blood-related diseases like Anaemia. In the medical image processing domain, precise pixel counting plays a significant role in diagnosing blood-related diseases. The proposed model is also outperformed in this regard. Table 3 shows the pixel count of healthy-RBCs and anaemicRBCs images individually, along with global pixel counting in a single blood image.

### D. COMPARATIVE ANALYSIS

A model proposed in [39] used 1,120 manually collected images for WBC detection without ground truth for authentication and pixel counting, which are key features for disease prediction using medical image processing. Authors in [44], Automated Semantic Segmentation of Red Blood

**TABLE 1.** Individual and global training, validation, test accuracies, BFScore, and IoU with relevant losses of healthy and anaemic-RBCs dataset.

| S # | Evaluation Matric | Healthy-RBC Dataset | Anaemic-RBC Dataset | Global |
|-----|-------------------|---------------------|---------------------|--------|
| 1 | Training Accuracy | 0.9856 | 0.9736 | 0.9796 |
| 2 | Training Loss | 0.0252 | 0.0351 | 0.0204 |
| 3 | Validation Accuracy | 0.9760 | 0.9696 | 0.9728 |
| 4 | Validation Loss | 0.0613 | 0.0776 | 0.0272 |
| 5 | Test Accuracy | 0.9720 | 0.9591 | 0.9655 |
| 6 | Test Loss | 0.0717 | 0.1085 | 0.0345 |
| 7 | BFScore | 0.9138 | 0.8978 | 0.9058 |
| 8 | IoU | 0.9311 | 0.9032 | 0.9171 |

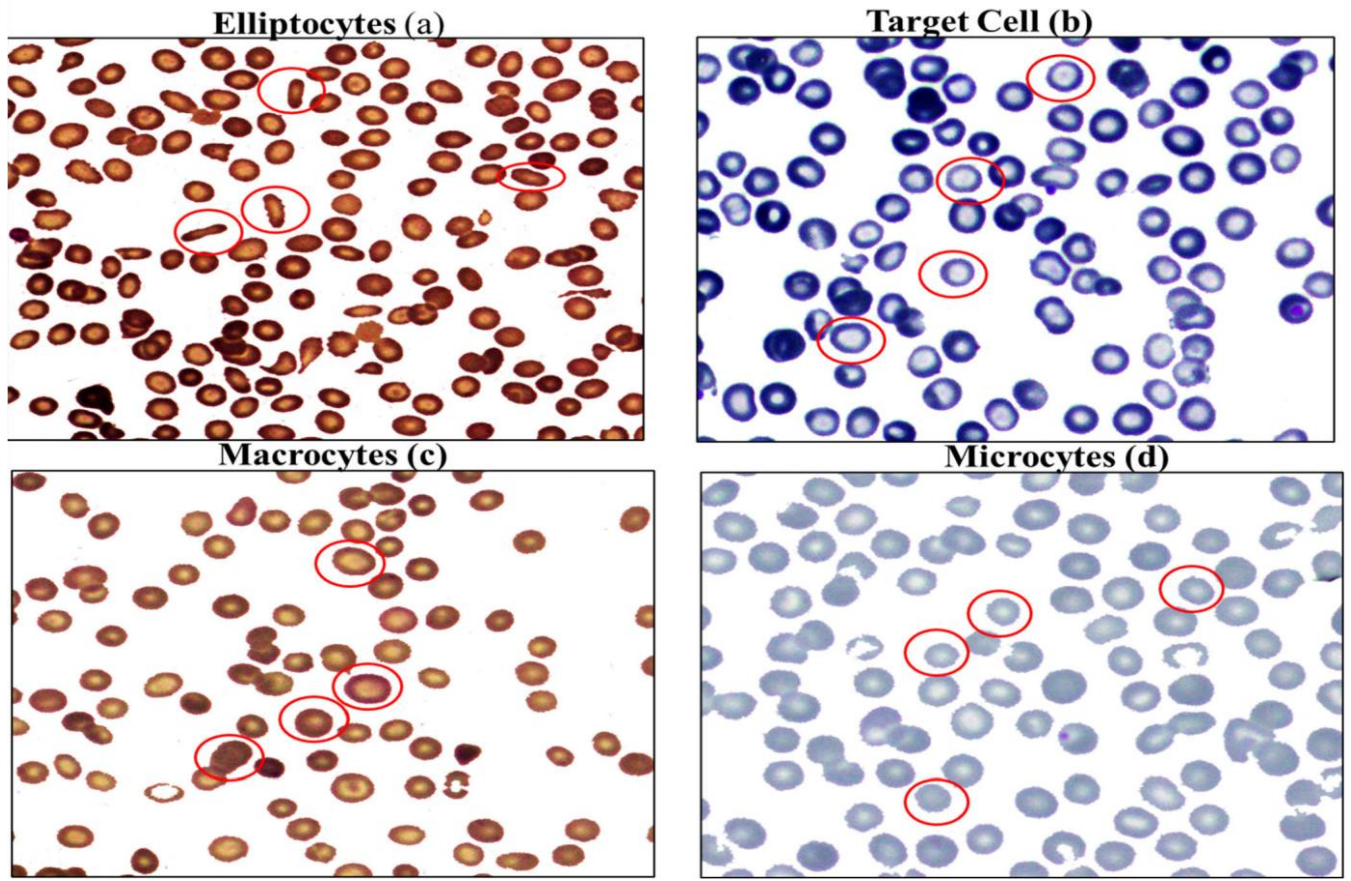

**FIGURE 8.** Shows the four categories of RBC elements, i.e., (a) Elliptocytes are elongated oval-shaped RBCs due to iron deficiency, (b) Target cells increase with excess cell membrane relative to cell volume, (c) Microcytes are those RBCs that have a size smaller than normal, and (d) Macrocytes RBCs are those whose size becomes larger than usual in healthy-RBC and Anaemic-RBC datasets.

Cells for Sickle Cell Disease target malaria and SCD without providing ground truth for crossmatching analysis. Research work carried out by [39], [42], [43], [71], [72] ignores the pixel counting, generation of ground truth, and targeting any disease. Most of the previous work ignores the training and validation accuracy with the highest accuracy of 98.87%, as shown in Table 4. This work first developed state-of-theart 2 RBC blood image datasets equipped with 1000 Healthy RBCs and 1000 anaemic-RBC images with 2000 ground truth for crossmatching analysis and performance evaluation. We have also categorized our dataset into four different RBC classes to classify the abnormal shape of the RBCs due to anaemia disease. To authenticate the proposed results, we have calculated three different types of accuracies along with relevant losses, i.e., training, validation, and testing accuracies and losses individually for both datasets as well as globally. The proposed model achieved individual training, validation, and testing accuracies of 98.56%, 97.60%, and 97.20%, respectively. While global accuracies were 97.96%, 97.28%, and 96.55% for training, validation, and testing. In addition to this, our model can also precisely count the pixels for diagnosing anaemia disease.

*E. ADVANTAGES OVER EXISTING TECHNIQUES*

This research work has upper hand over current research work relevant to segmentation of RBC image in respect of the following:

1) We used a multi-level deep convolutional encoderdecoder network that preserved the semantic information of the blood element.

2) Each level of the network performs a segmentation training process upto a specific threshold value that helps to reduce unnecessary training.

3) We have developed two state-of-the-art RBCs datasets with 2000 original images (1000 healthy and 1000 anaemic) and 2000 manually generated ground truth with approximately 1,992,000 RBC elements for the research community.

4) Each level has its specific learning weights that help to extract feature maps upto a specific level. So the network is prevented from the extraction of unnecessary features.

5) The multi-level approach also reduces the gradient descent phenomenon during the training.

6) The chances of over-fitting and under-fitting of the network become negligible using $T_o$ value at each level.

TABLE 2. Dataset comparison with proposed novel RBCs datasets. "X" represents non-availability and "✓" availability.

| Dataset | No of Images | Blood elements | Ground truth | CBC reports | Morphology reports |
|---|---|---|---|---|---|
| ALL-IDB-I and II [65] | 108 and 260 | WBC and RBC | × | × | × |
| SMC-IDB[68] | 367 | WBC | × | × | × |
| IUMS-IDB [67] | 196 | WBC | × | × | × |
| Malarial dataset | 848 | RBC | × | × | × |
| **Proposed Healthy-RBC Dataset** | **2000** | **RBC** | ✓ | ✓ | ✓ |
| **Proposed Anaemic-RBC Dataset** | **2000** | **RBC** | ✓ | ✓ | ✓ |

TABLE 3. Pixel count of healthy and anaemic-RBCs images along with global pixel count.

| S # | Blood cell image Type | Pixel count |
|---|---|---|
| 1 | Healthy-RBCs Image | $3.2657e + 09$ |
| 2 | Anaemic-RBC Image | $2.1728e + 08$ |
| 3 | Global pixel count | $1.7414e + 09$ |

TABLE 4. Brief comparative analysis of proposed dataset and technique with current research work.

| | Dataset Comparisons | | | | | Technique evaluation comparisons | | |
|---|---|---|---|---|---|---|---|---|
| Work | Dataset | No of images | Ground Truth | Blood element | Pixel counting | Targeted Disease | Accuracy type | Evaluation matrices | Accuracy achieved |
| [39] | Manually collected | 1,120 | × | WBC | × | × | Mean accuracy | Accuracy | 90.09% |
| [44] | Manually collected | 848 and Malarial Dataset | × | RBC | × | Malaria | Global accuracy | Accuracy | 93.72% |
| [66] | Extended ALL-IDB-I, II | Original 108, Ground truth 108 | RBC, WBC, plt | ✓ | × | Mean, global Accuracy | Accuracy, IoU and BFScore | 97.18% |
| [71] | Manually collected | 564 | × | WBC | × | × | Accuracy | Precision, recall and specificity | 100% |
| [72] | SMC-IDB, IUMS-IDB, ALL-IDB | 367, 195 and 108 | × | WBC | × | × | Accuracy | Accuracy | 94.1% |
| [40] | Manually collected | 314 | × | RBC | ✓ | SCD | Not mentioned | Dice Coefficient, Jaccard Index, and Hausdorff Distance | Not mentioned |
| [42] | ALL-IDB-I and II | 42 and 101 | 42 | WBC and RBC | × | × | Accuracy | Global accuracy | 93.3% |
| [43] | Manually collected | 257 | × | WBC | × | × | Accuracy | Accuracy | 98.87 |
| **Proposed** | **Healthy-RBCs and Anaemic-RBCs** | **2000 and 2000** | ✓ | **RBC** | ✓ | **Anaemia** | **Training Accuracy, Validation accuracy, Test Accuracy and Global accuracy** | **Accuracy, IoU and BFScore** | **Global Training = 97.96%, Validation = 97.28%, Test = 96.55%** |

7) We have performed 5-fold training, validation, and testing of blood segmentation to authenticate the proposed results.

8) We also compare the proposed results with manually generated ground truth and evaluate it by finding IoU. Our model got 0.9311and 0.9032 for the healthy and anaemic-RBC dataset with a MeanIoU of 0.9171 on both datasets.

## VI. CONCLUSION

To handle the precise segmentation of blood elements in digital images, we proposed a novel model known as multi-level DCED-Net. The multi-level concept enhances the preservation of semantic information at the pixel level, which helps for accurate, effective segmentation. For experimental purposes, we have developed two state-of-the-art RBC datasets for the research community. These datasets will play a pivotal role in determining an accurate model in the semantic segmentation domain and diagnosing anaemia deformities in RBC cells. During the experiment, we carried out 5-fold training, validation, and testing to obtain precise results. Our model got 0.9856, 0.9760, and 0.9720 training, validation, and testing accuracies on the healthy-RBC dataset, respectively. The anaemic-RBC dataset, training, validation, and testing accuracy were 0.9736, 0.9696, and 0.9591, respectively. We have generated a total of 2000 ground truth mask images for crossmatching analysis and got the IoU 0.9311 and 0.9032 with BFScore 0.9138, 0.8978 on healthy and anaemic datasets.

## FUTURE WORK

In the future, we will enhance our model for pixel-level classification regarding RBCs classes, i.e., normal, microcytes, macrocytes, elliptocytes, and target RBCs. This work is the first phase of the Anaemia Disease Prediction Support System. The next paper will extend this dataset and classify each RBC type, i.e., normal, microcytes, macrocytes, elliptocytes, and target RBCs. This classification will help to detect anaemia disease automatically in blood cells.

## FUNDING

No funding is available for this study

## AUTHORSHIP CONTRIBUTION

**Muhammad Shahzad:** Conceptualization; Methodology; Formal analysis; Data curation; Writing - Original Draft; Writing - review & editing, Dataset Preparation. **Arif Iqbal Umar**: Supervision and reviewing. **Syed Hamad Shirazi**: Reviewing, Editing, Data Analysis. **Israr Ahmed Shaikh**: Image Collection, Image Preparation.

## COMPETING INTEREST

All authors declare no conflict of interest.

## AUTHORS APPROVAL

All authors have read and approved the final manuscript.

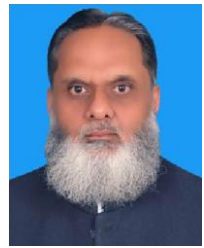

ARIF IQBAL UMAR is an Associate professor in the Department of Computer Science & Information Technology Mansehra, Pakistan. He obtained his MSc (Computer Science) degree from the University of Peshawar, Peshawar, Pakistan, and Ph.D. (Computer Science) degree from BeiHang University (BUAA), Beijing P.R. China. His research interests include DataMining, Machine Learning, Information Retrieval, Digital Image Processing, Computer Networks Security, and Sensor Networks. He has at his credit 22 years' experience of teaching, research, planning, and academic management.

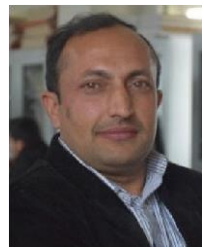

SYED HAMAD SHIRAZI is an Assistant professor in the Department of Computer Science & Information Technology, Hazara University Mansehra, Pakistan. He received his Ph.D. degree from the Department of Information Technology, Hazara University Mansehra. He has achieved his MS Computer Science degree from COMSATS Abbottabad, Pakistan. His skills and expertise are Computer Vision, Texture analysis, Neural networks, Object recognition, Pattern Recognition, Digital Image processing, and machine learning and wavelet transformation.

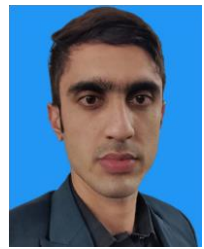

ISRAR AHMED SHAIKH is currently working as a Pathologist at Shaukat Khanum Memorial Cancer Hospital and Research Centre, Lahore, Pakistan. His research interests include disease diagnostic or therapeutic, blood cell analysis, clinical diagnostic testing pathology, and laboratory medicine services.

• • •

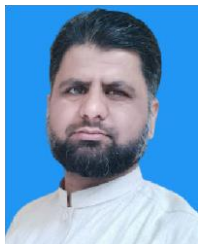

MUHAMMAD SHAHZAD received the M.S. degree in computer science from the Virtual University of Pakistan. He is currently a Ph.D. scholar with the Department of Information Technology, Hazara University Manshera, Pakistan. His research interests include data mining, machine learning, deep learning, medical image processing, and bioinformatics.